**Current Controlled Magnetization Switching in Cylindrical Nanowires for High-Density 3D Memory Applications**

*Hanan Mohammed[1], Hector Corte-León[2], Yurii P. Ivanov [1,3,4], Sergei Lopatin[5], Julian A. Moreno[6], Andrey Chuvilin[7,8], Akshaykumar Salimath[6], Aurelien Manchon[1,6], Olga Kazakova[2], and Jurgen Kosel[1] \**

[1]King Abdullah University of Science and Technology, Computer Electrical and Mathematical Science and Engineering, Thuwal, 23955-6900, Saudi Arabia

[2]National Physical Laboratory, Teddington, Hampton Road, TW11 0LW, United Kingdom

[3]Erich Schmid Institute of Materials Science, Austrian Academy of Sciences, Jahnstrasse 12, A-8700, Leoben, Austria
[4]School of Natural Sciences, Far Eastern Federal University, 690950, Vladivostok, Russia

[5]Imaging and Characterization Core Lab, King Abdullah University of Science and Technology (KAUST), Thuwal 23955-6900, Saudi Arabia.

[6]King Abdullah University of Science and Technology, Physical Science and Engineering, Thuwal, 23955-6900, Saudi Arabia

[7]CIC nanoGUNE Consolider, Avenida de Tolosa 76, 20018 San Sebastian, Spain
[8]IKERBASQUE, Basque Foundation for Science, Maria Diaz de Haro 3, 48013 Bilbao, Spain

e-mail: jurgen.kosel@kaust.edu.sa.

A next-generation memory device utilizing a three-dimensional nanowire system requires the reliable control of domain wall motion. In this letter, domain walls are studied in cylindrical nanowires consisting of alternating segments of cobalt and nickel. The material interfaces acting as domain wall pinning sites, are utilized, in combination with current pulses, to control the position of the domain wall, which is monitored using magnetoresistance measurements. Magnetic force microscopy results further confirm the occurrence of current assisted domain wall depinning. Data bits are therefore shifted along the nanowire by sequentially pinning and depinning a domain wall between successive interfaces, a requirement necessary for race-track type memory devices. We demonstrate that the direction, amplitude and duration of the applied current pulses determine the propagation of the domain wall across pinning sites. These results demonstrate a multi-bit cylindrical nanowire device, utilizing current assisted data manipulation. The prospect of sequential pinning and depinning in these nanowires allows the bit density to increase by several Tbs, depending on the number of segments within these nanowires.

Next generation data storage concepts are governed by the need for high density, low cost and fast operation. The racetrack memory is such a concept, wherein data is stored magnetically as a series of magnetic bits along a nanowire.[1] For read/write operations, the bits are moved along the nanowire using a spin polarized current.[2-3] The proposed concept has been widely explored using lithographically patterned 2D nanowires.[4-7] While significant advances have been made in such planar devices, the envisioned 3D architecture has not been achieved, due to the complexity of its fabrication. An attractive alternative to 2D planar nanowires are cylindrical nanowires, which are grown inside of anodic alumina oxide (AAO) templates.[8-11] This low-cost fabrication technique allows a precise control over nanowire density, spacing, diameter and length.[12] The shape anisotropy property of these nanowires generally results in a single magnetic domain structure within them, implying a single bit per nanowire.[13] It has been shown that pinning potentials for domain walls (DWs) can be created by modulations in diameter [14-17] and chemical composition.[18-20] With these methods, multiple DWs can be introduced, thereby increasing the number of domains or bits per single nanowire. With the storage capacity of a cylindrical nanowire device depending on factors such as number of nanowires, number of bits per nanowire and length of nanowire, data storage densities can be tuned by several folds.

Recent studies on cylindrical nanowires of certain diameters have confirmed the existence of a unique type of topologically protected DW, the Bloch-point domain wall.[19, 21] This type of DW consists of a magnetic 'hedgehog' structure around a Bloch-point, which is a point of vanishing magnetization.[22-26] From theoretical studies, Bloch-point domain walls are associated with an ultra-high propagation speed in the range of 1 km s$^{-1}$,[27] which is a very attractive feature for data storage applications. The multisegmented Co/Ni nanowires of 80 nm diameter utilized in this research have been reported to have a 3D spin structure with a Bloch-point at its center.[28] By shifting the DWs across multisegmented nanowires using current pulses, a fast 3D memory device could be realized. Thereby with a nanowire density in the order of 10$^{11}$ nanowires cm$^{-2}$ in AAO templates,[12] a storage capacity of 50 Tb inch$^{-2}$ and even higher could be achieved. Although studies have demonstrated magnetic field assisted motion of a DW in cylindrical nanowires,[29] insight into current assisted DW depinning has been limited due to several challenges.[28] The current assisted DW motion previously shown utilizes a DC current in the presence of a background field to depin a domain wall.[28] In this case, the DC current allowed changing the configuration from a two-domain nanowire to a single domain nanowire, effectively realizing a single bit device. Successive pinning and depinning could not be shown. Pulsed-current operation is therefore essential for a multi-bit device, in order to shift DWs across individual interfaces.

In this research, we utilize current pulses in the presence of a background magnetic field to manipulate DWs between pinning sites. Pinned DWs are visualized in multisegmented Co/Ni nanowires using magnetic tomography. Using magnetoresistance measurements and magnetic force microscopy techniques, we demonstrate for the first time, the movement of the DW between successive pinning sites, thereby demonstrating a two-bit nanowire system consisting of four states. We observe that the direction, amplitude and duration of the applied current pulses determine the depinning and propagation of the DW. These results demonstrate a scalable technology based on cylindrical nanowires for multi-bit 3D memory devices.

The multisegmented Co/Ni nanowires were grown by electrodeposition into Anodic Aluminum Oxide templates followed by chemical release of the nanowires from the template, resulting in nanowires with a diameter of 80 nm and lengths of 30 µm.[9] Magnetic tomography measurements were performed to reconstruct the 3D spin structure of the nanowire, utilizing a novel Virtual Bright Field Differential Phase Contrast (VBF-DPC) imaging technique.[30]

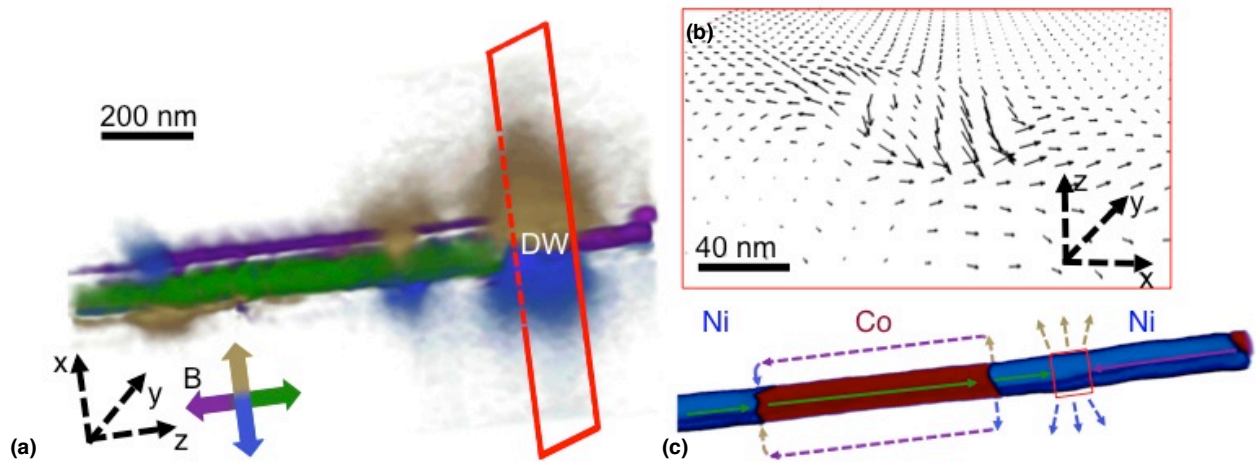

**Figure 1. Transmission electron microscopy analyses of a multisegmented Co/Ni nanowire: (a)** magnetic tomography depicting 3D distribution of the magnetic field (*B*). Green indicates magnetization to the right, whereas purple indicates magnetization to the left. Brown and blue indicate up and down magnetization, respectively. A domain wall (DW) separates opposing magnetization directions. **(b)** An arrow plot of the spin distribution at the nanowire cross-section extracted from (a) reveals a head-to-head domain wall with a 3D distribution of spins and **(c)** schematic of 3D distribution of the magnetic field overlapped on a compositional tomography of the nanowire. The solid arrows indicate magnetization of the nanowire, whereas the stray magnetic field is indicated by the dashed arrows. The red square indicates the position of the DW.

Figure 1 (a) depicts the 3D distribution of the magnetic field in the Co/Ni nanowire. Figure 1 (c) depicts the Electron Energy Loss Spectroscopy (EELS) elemental mapping of the nanowire with an overlapping schematic of the magnetic field. The Ni segment on the left end of the nanowire and the adjacent Co segment are magnetized from left to right (green contrast within the nanowire in Figure 1 (a)), whereas a major portion of the Ni segment on the right end of the nanowire is magnetized in the opposite direction (purple contrast within the nanowire). Due to the larger magnetization value of Co compared to the Ni segment, stray fields can be observed at the interfaces, (blue and brown contrast). In addition to the stray field from the interfaces, a larger stray field is observed, as a result of a pinned DW (red rectangle in Figure 1 (a)). The position of the DW away from the interface is a consequence of the larger magnetization of the Co segment.[19] A detailed analysis of the DW from Figure 1 (b) reveals a 3D distribution of spins shown in Figure 1 (b).

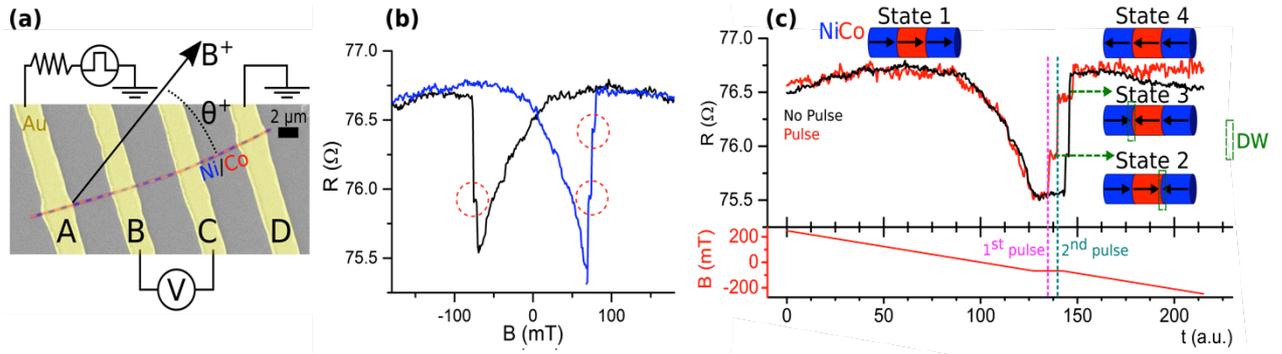

**Figure 2. Magnetoresistance (MR) measurements of a single multisegmented Co/Ni nanowire.** (a) Electrical setup and nanowire composition extracted from energy dispersive x-ray imaging. (b) Individual MR curves, with field applied at $\theta = 63°$ to the nanowire. Black curve: field swept from positive to negative; Blue curve: field swept from negative to positive; Red circles indicate domain wall pinning events. (c) Temporal evolution of MR (top) and magnetic field magnitude (bottom) when field is applied at $\theta = 63°$ to the nanowire. Black curve: no current pulse is applied to the nanowire; red curve: current pulses are applied to the nanowire (1st and 2nd pulse respectively). The nanowire schematics in (c) represent the magnetization direction at specific points in the MR evolution thereby depicting a two-bit system consisting of four states.

Anisotropic magnetoresistance (MR) measurements are used to gain an insight into the magnetization reversal process in individual nanowires (see Supporting Information for Single Nanowire Device Fabrication). The MR value depends on the relative orientation of the magnetic moments with respect to the applied elecrtric current.[9] In this study, MR measurements were performed at room temperature to monitor the nanowire's magnetization reversal and determine the pinning and depinning of DWs. The electrical setup and device structure (as extracted from Energy Dispersive X-ray (EDX) imaging) is shown schematically in Figure 2 (a). We first determine the switching field of the nanowire by passing a DC current of 100 µA (current density $J \sim 2 \times 10^{10}$ A m$^{-2}$) through the outermost electrodes (from A to D in Figure 2 (a)) and measure the voltage between the inner electrodes (B and C), while an external magnetic field (*B*) was ramped up and down. The external field was applied at an angle $\theta = 63°$ with respect to the nanowire. Orienting the magnetic field at an angle, instead of parallel to the nanowire increases the number of DW pinning events, facilitating the study of DW pinning/depinning mechanism.[31] Figure 2 (b), depicts the typical field dependence of MR (blue/black curves for field ramping up/down respectively), where the high resistance state is a consequence of the magnetic moments within the nanowire being aligned parallel to the direction of the current. Following the black curve in Figure 2 (b) (i.e. field ramping up), increasing the field from negative saturation towards positive values leads to a rotation of the magnetic moments away from the direction of current, resulting in a gradual reduction of resistance. At ~70 mT, the resistance of the nanowire increases abruptly in three successive steps. Thus, a DW has been nucleated at ~70 mT, which then propagates along the nanowire and gets pinned at a Co/Ni interface, thereby reversing the magnetization only within a section of the nanowire. Upon further increase in the magnetic field, the DW gets depinned (at ~78 mT), propagates along the nanowire and then pins at a subsequent pinning site. Applying an even higher magnitude of magnetic field (~86 mT) depins the DW from the second pinning site, leading to a complete magnetization reversal as reflected by the final high resistance state in the MR curve. The first and second pinning events are marked by the red circles in Figure 2 (b). Ramping the field down (i.e. red curve), leads to similar results, but in this case, due to pinning/depinning stochasticity, only two steps can be seen.

Prior to investigating the influence of current pulses on a DW, a control MR measurement was performed ramping the field from positive to negative values (black curve in Figure 2 (c)). The procedure followed was the same as in the previous MR measurement except that in this case, before reaching the nanowire's switching field, the external field was kept constant at -68 mT for a period of 30 s (see bottom panel in Figure 2 (c)). Then, the field was continuously decreased until the direction of magnetization within the entire nanowire was reversed. Here, we observe no variation in resistance when the external field is constant (i.e. at 68 mT). From this control MR measurement, we conclude that the DW propagates only upon increasing the external field above the switching field; thus, allowing us to disregard any stochastic depinning effects that could occur in the nanowire, when applying pulses of current.

In order to study the influence of current pulses on a pinned DW, the same procedure stated previously was followed except that now in addition to the constant external field of -68 mT, a current pulse (1st pulse) was injected from electrode A to D ($I_{DC}$ = 4 mA, $t_{pulse}$ = 600 ns with a pulse rise time of ~ 8 ns), while still monitoring the resistance (red curve in top panel of Figure 2 (c)). The current pulse was applied opposite to the direction of the external field (*B*), i.e. the electron flow and *B* were in the same direction. After a duration of 10 s, a second pulse (2nd pulse) of the same amplitude and duration as the 1st pulse was applied. The timings of the application of pulses are indicated by the dashed vertical lines. The application of the first pulse resulted in an immediate increase in the resistance; however, it should be noted that the nanowire has not yet reached its highest resistance value. This indicates the presence of a pinned DW within the nanowire. Furthermore, the application of the second pulse also resulted in an immediate increase of resistance. This once more indicates depinning of the DW from its previous position and its consequent propagation until it gets pinned at the next pinning site. Further decreasing the magnetic field towards negative saturation values result in the depinning and propagation of the DW leading to the complete magnetization reversal in the nanowire solely under the influence of the external field.

To confirm whether the observed DW motion is a consequence of the applied pulses (i.e. 1st and 2nd pulses) and not the external field and/or heating effects, a reverse orientation of the field was applied. Note that the direction of the applied current pulses remained unchanged. The results (shown in Supporting Information), demonstrate no change in resistance after the application of the 1st and 2nd pulses. This indicates that the mutual orientation of *I* and *B* are crucial for the DW depinning and propagation process. This also confirms that DW motion was triggered by the current pulse and rules out depinning due to heating effects, which are small, as has been previously shown.[28] Taking into consideration the direction of saturation of the nanowire, the external magnetic field, the direction of the applied current and the occurrence of DW motion, we conclude that a head-to-head DW moved from the right end of the nanowire to its left end. Thus, the combination of the external field and electron flow in the same direction (left), led to the propagation of the head-to-head DW towards the left end of the nanowire (see schematics in Supporting Information).

The observation that a head-to-head DW can be depinned and moved by a current pulse, when the directions of electron flow and external field are in the same direction was further confirmed by applying pulses with a varying range of pulse amplitude and duration (**Figure 3**). The application of a wide range of pulse amplitudes (up to 2 mA) and durations (up to 1500 ns) allows to evaluate the current density required for DW depinning and whether the depinned DW could get pinned again at a subsequent interface.

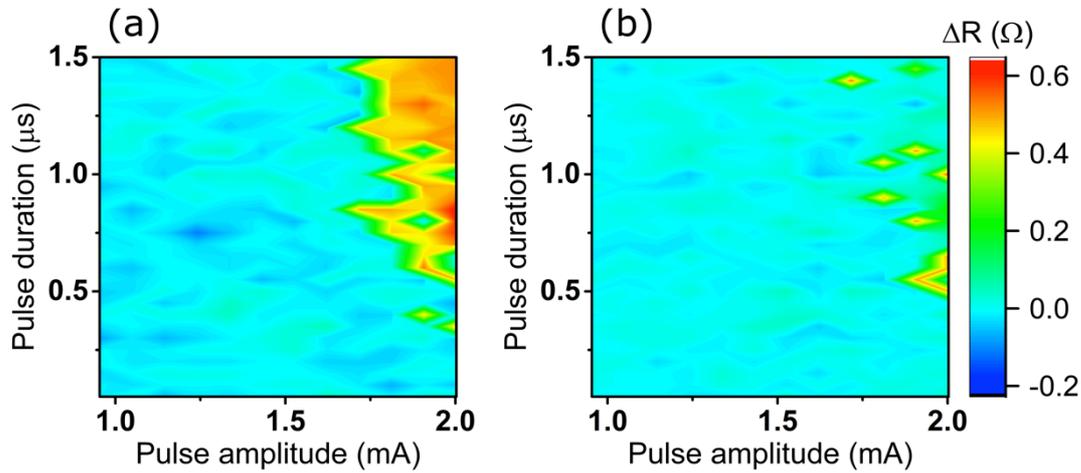

**Figure 3. Change in the resistance ($\Delta R$) obtained from magnetoresistance measurements of a multisegmented Co/Ni nanowire by applying current pulses of varying amplitude and duration. (a)** Application of current (1st pulse), **(b)** current (2nd pulse). The electron flow and external field ($B$) are in the same directions.

The measurement procedure is the same as the one in Figure 2 (c). The change in resistance ($\Delta R$) is measured with respect to the resistance immediately before applying the current pulse. Figures 3 (a) and (b) depict the influence of 1st and 2nd pulses, respectively, when the applied external field and electron flow are in the same direction. The results show that a change in resistance does not occur at low pulse amplitudes (up to 1.5 mA), indicating that the DWs are not affected by the applied pulse (Figure 3 (a)). At higher pulse amplitudes and lower pulse durations, a change in resistance is observed (depicted by the green regions in Figure 3 (a)), indicating the reversal of magnetization within a section of the nanowire. This implies that the DW propagates a short distance, as it gets pinned at a subsequent interface, at a current density of $J \sim 3 \times 10^{11}$ A m$^{-2}$. A higher pulse amplitude and longer duration resulted in a larger $\Delta R$ (depicted by yellow and red regions in Figure 3 (a)), indicating that the magnetization within a larger section of the nanowire has reversed. Thus, the DW has propagated a larger distance and completely reversed the magnetization direction within the nanowire. We can thus conclude that the higher current densities move the DW across multiple pinning sites thereby enabling multi-bit device applications. Figure 3 (b) depicts the effect of 2$^{nd}$ current pulse on the magnetization state of the nanowire. The 2$^{nd}$ pulse affected the DWs that had previously been pinned at an interface (depicted by the yellow regions in Figure 3 (a)) and succeeds in depinning the DWs, thereby reversing the magnetization within the nanowire. These measurements are in agreement with our previous findings and indicate that the direction of the head-to-head DW propagation is that imparted by the electron flow, in the direction of the external field.

In addition to MR measurements, *in situ* Magnetic Force Microscopy (MFM)[32] was employed to observe the effect of current pulses on pinned DWs. A multisegmented Co/Ni nanowire with electrical contacts, was first saturated at $B$ = -80 mT (State 1 in Figure 4, with the external field applied horizontally i.e. $\theta$ = 0°). Periodic bright and dark contrasts can be observed, which are a consequence of the strong stray fields emanating from the Co segments. These contrasts aid in understanding the magnetization within the nanowire, since a reversed periodic contrast (i.e. dark followed by bright) would indicate a reversal of magnetization within the Co segment.

In order to nucleate a DW, the external field was increased, and at 50 mT (State 2 in Figure 4), the stray field contrasts from two Co segments (center and right) have a reversed magnetization. Thus, a DW has been nucleated at the right end of the nanowire, which then propagates and gets pinned in the center of the nanowire. Considering the direction of saturation of the nanowire, the externally applied field and propagation of the DW, we can conclude that a tail-to-tail DW is pinned at one of the interfaces. To investigate the effect of current pulses on the pinned tail-to-tail DW, a current pulse (pulse amplitude of 4 mA and pulse duration of 600 ns) was applied at State 2. The application of the current pulse in the same direction as the external field leads to the reversal of magnetization within the left Co segment (as seen by the reversal of the Co stray field contrast as well as the segment outside electrode B, in State 3 of Figure 4). Thus, the tail-to-tail DW moves in the direction of the electron flow, i.e. towards the left end of the nanowire, and the opposing direction of $B$ assists its propagation. In addition to this, a control measurement was performed, where instead of applying a pulse at State 2, the external field was increased until the nanowire reversed its magnetization at $B = +80$ mT (State 3' in Figure 4). We observe that states 3 and 3' are equivalent magnetization states. Thus, we demonstrate the depinning of a tail-to-tail DW upon the application of a current pulse in the presence of a background magnetic field provided that the direction of $B$ assists in DW motion. The MFM results further confirm that the change in resistance observed in the MR upon the application of a current pulse, is in fact the movement of a DW (as evidenced by the switching of the Co segment) and not merely rearrangement of the magnetization within the nanowire.

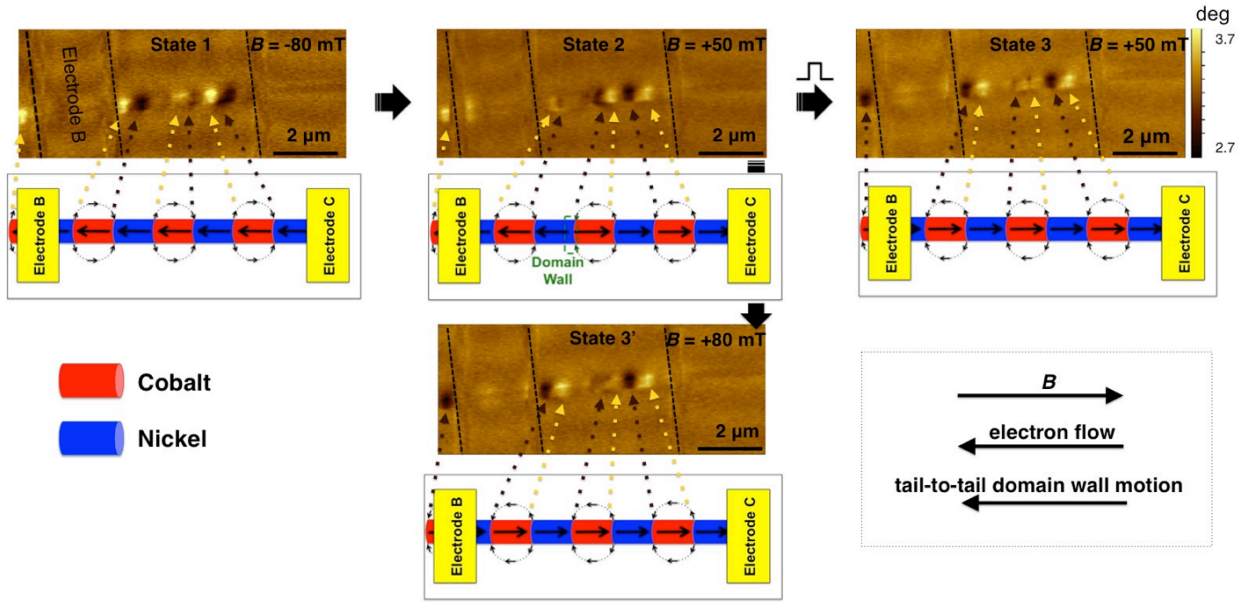

**Figure 4. Magnetic Force Microscopy (MFM) images (top) and corresponding nanowire schematics (bottom) of a multisegmented Co/Ni nanowire.** State 1 corresponds to the saturated state when magnetic field is applied in $-B$ direction (-80 mT). The black arrows within the nanowire schematics indicate the direction of magnetization and the black arrows outside the Co segment represent the stray fields emanating from the Co segment. State 2 corresponds to $B = 50$ mT where a tail-to-tail DW has been nucleated. A current pulse (of the same direction as $B$) was applied to state 2 resulting in the propagation of the tail-to-tail DW towards electrode B. State 3' corresponds to the saturated state in $+ B$ direction (at 80 mT) obtained by increasing the external field in state 2, without the application of a current pulse.

In this letter, we present for the first time an experimental observation of sequential pinning and depinning of DWs, which is essential for memory devices in cylindrical nanowires, enabled by using current pulses. We utilized the interfaces of multisegmented cylindrical Co/Ni nanowires to act as pinning sites and demonstrated the motion of DWs between successive pinning sites, thereby demonstrating a multi-bit memory device based on cylindrical nanowires. Additionally, we demonstrated that the combination of amplitude and duration of applied pulses determine a complete parametric space where the DW can be depinned and propagated across multiple pinning sites with a minimum current density of $J \sim 3 \times 10^{11}$ A m$^{-2}$. Head-to-head DW motion was shown using magnetoresistance measurements, where the external magnetic field and electron flow were oriented in the same direction, whereas opposing orientation of the magnetic field and electron flow hindered its motion (Figure 2, 3 and S1). On the other hand, the motion of tail-to-tail DW required opposing orientations of external magnetic field and electron flow, which has been shown using magnetic force microscopy measurements (Figure 4). The four states that were observed in the nanowire represent a two-bit system, demonstrating the concept of a cylindrical nanowire memory that is operated by current pulses. The ability to sequentially pin and depin a DW along the nanowire opens up the possibility to introduce several bits per nanowire. Such a device could be scaled to obtain fast 3D memories with ultra-high data storage density of over 50 Tb inch$^{-2}$.

**ASSOCIATED CONTENT**

**Supporting Information**

Supporting Information is available free of charge

Methods, experimental setup and additional magnetoresistance measurement

## ACKNOWLEDGMENTS

This work was funded partly by King Abdullah University of Science and Technology (KAUST) and in part by EMRP and EMRP participating countries under Project EXL04 (SpinCal), and FP7 project NanoMag, and NanoMag, MagNaStand (EMPIR). The image for the graphical table of content was produced by Xavier Pita, scientific illustrator at King Abdullah University of Science and Technology (KAUST).